\documentclass[A4,12pt]{article}

\makeatletter
 
\@addtoreset{equation}{section}
\makeatother

\newcommand{\hs}{\hspace}
\newcommand{\ts}{\hspace*}
\newcommand{\vs}{\vspace}
\newcommand{\e}{\enskip}
\newcommand{\q}{\quad}

\newcommand{\dps}{\displaystyle}
\newcommand{\f}{\frac}

\newcommand{\Phat}{\hat{{\cal P}}}
\newcommand{\PhatI}{\hat{{\cal P}}^{\mbtn{(1)}}}
\newcommand{\Qhat}{\hat{\Omega}}

\newcommand{\Zp}{\hat{Z}^{(+)}}
\newcommand{\Zm}{\hat{Z}^{(-)}}

\newcommand{\ximm}[1]{\hat{\xi}^{\mbtn{#1}(-)}}

\newcommand{\pimm}[1]{\hat{\pi}^{\mbtn{#1}(-)}}

\newcommand{\ximms}[2]{\hat{\xi}^{\mbtn{#1}(-)}_{#2}}

\newcommand{\pimms}[2]{\hat{\pi}^{\mbtn{#1}(-)}_{#2}}

\newcommand{\Zmms}[2]{\hat{Z}^{\mbtn{#1}(-)}_{#2}}

\newcommand{\bM}{\bar{M}}
\newcommand{\mcK}{\mathcal{K}}
\newcommand{\mcC}{\mathcal{C}}
\newcommand{\mcA}{\mathcal{A}}

\newcommand{\mcS}{\mathcal{S}}

\newcommand{\mcG}{\mathcal{G}}

\newcommand{\mcD}{\mathcal{D}}
\newcommand{\mcU}{\mathcal{U}}

\newcommand{\mcW}{\mathcal{W}}

\newcommand{\commut}[2]{[#1,\e #2]}
\newcommand{\symmp}[2]{\ \{#1,\e #2\}}
 
\newcommand{\mbtn}[1]{\mbox{{\tiny #1}}}
\newcommand{\pscrp}{\mbox{{\scriptsize $\Phat$}}} 
\newcommand{\starp}{\mbox{{\scriptsize š}}}

\newcommand{\inM}{M^{-1}}

\begin{document}

\begin{center}
{\bf \Large Noncommutative Quantum Mechanics on a Curved Space
}\vs{12pt}\\
M. Nakamura\footnote{\label{*}E-mail:mnakamur@hm.tokoha-u.ac.jp}
\vs{12pt}\\
{\it Research Institute, Hamamatsu Campus, Tokoha University, Miyakoda-cho 1230, 
Kita-ku, Hamamastu-shi, Shizuoka 431-2102, Japan}
\end{center}

\begin{abstract}
Starting with the first-order singular Lagrangian, the canonical structures of the noncommutative quantum system on a submanifold embedded in the higher-dimensional Euclidean space are investigated with the projection operator method (POM) and the Dirac-bracket formulation in the case of the derivative-type constraint. Using the successive projection procedure and the iterativity of the Dirac bracket, the noncommutative quantum system is constructed in the form including all orders of the noncommutativity-parameters. When the noncommutative quantum system is constrained to a curved space, the commutator algebra of the system is presented within the 1st-order approximation with respect to $\hbar$ and the noncommutativity-parameters. 
\end{abstract}


\section{Introduction}	

\ts{12pt}The problem of the quantization of a dynamical system constrained to a submanifold   embedded in the higher-dimensional Euclidean space has been extensively investigated as one of the quantum theories on a curved space until now\cite{A1,A2,A3,A4,A5}. In order to avoid the unnecessary troublesomeness, the submanifold $M^{N-1}$ specified by $G(x)=0$ ($G(x)\in {\it C}^{\infty}$) in an $N$-dimensional Euclidean space $R^N$ has been considered in many studies, where $x=(x^1,\cdots,x^i,\cdots,x^N)\in R^N$. \\
\ts{12pt}For the dynamical system constrained to $M^{N-1}$, there have been studied the two kinds of constraint conditons, one of which is the {\it static} constraint, $G(x)=0$, and the others, the {\it dynamical} one, $\dot{G}(x)=0$\cite{A2}. Then, it has been shown that the {\it static} constraint yields the noncommutativity among the canonically conjugate operators\cite{A1,A6}, and the {\it dynamical} one conserves the canonically conjugate commutation relations on the flat space\cite{A2,A4,A5,A6}. Therefore, it is extremely interesting to extend the quantum mechanics on a curved space to the noncommutative quantum mechanics\cite{A7}. \\
\ts{12pt}Following our previous works\cite{A2,A8}, in this paper, we shall investigate the  noncommutative quantum system on a curved space in the framework of the operatorial quantization formalism for constrained systems\cite{A9}. \\
\ts{12pt}Starting with the Faddeev-Jackiw type\cite{A10} first order singular Lagrangian containing the term associated to the {\it dynamical} constraint, we shall construct the canonical structure of the noncommutative quantum system exactly through the projection operator method(POM) with the constraint star-products\cite{A11}. Then, it will been shown that the commutator algebra of the resultant system and the Hamiltonian hold the quantum corection terms due to the noncommutativity associated to the constraint opetator $G(x)$. Such a noncommutativity will yield the complicated structure in the commutator algebra on a curved space and the projection of Hamiltonian. Then, in this paper, the canonical structure on a curved space will be constructed within the 1st-order approximation with respect to both of $\hbar$, that is, the Dirac-bracket quantization\cite{A12}, and  noncommutativity-parameters. \\
\ts{12pt}This paper is organized as follows. In Sect.2, we propose the Lagrangian with the {\it dynamical} constraint and construct the initial unconstraint quantum system, which we denote $\mcS$. In Sect.3, we construct the the resultant constraint quantum system, which we denote $\mcS^*$. In Sect.4, the discussion and the some concluding remarks are given. 

\section{Initial Hamiltonian System}

\ts{12pt}We here propose the first-order singular model Lagrangian describing the noncommutative quantum system constrained to the curved space. Following the canonical quantization formulation for constraint systems\cite{A8,A12}, then, we shall construct the unconstraint quantum system $\mcS$.

\subsection{Noncommutativity Matrix $\Theta$, $\Xi$}

\ts{12pt}Let $\Theta$ and $\Xi$ be the totally antisymmetric matrices defined as follows:  
$$
\Theta=\theta\varepsilon,\hs{48pt}\Xi=\eta\varepsilon,
\eqno{(2.1)}
$$
where $\theta$ is the constant parameter describing the noncommutativity of coordinates and $\eta$, that of momenta, and
$\varepsilon$ is the completely antisymmetric tensor defined as
$$
\varepsilon^{ij}=1\hs{12pt}(i>j), \hs{12pt}\varepsilon^{ji}=-\varepsilon^{ij}\hs{24pt}(i,j=1,\cdots,N).
\eqno{(2.2)}
$$
These matrices satisfy 
$$
\Theta\Xi=\Xi\Theta,\hs{48pt}(\Theta\Xi)^t=\Theta\Xi.
\eqno{(2.3)}
$$
In terms of $\Theta$ and $\Xi$, then, the following matrices are defined:
$$
\begin{array}{l}
G=\Theta\Xi=\Xi\Theta,\vs{6pt}\\
M=I+\dps{\f14}G,\hs{24pt}\bM=I-\dps{\f14}G,
\end{array}
\eqno{(2.4)}
$$
which are symmetric and commutable with $\Theta$, $\Xi$, and therefore become commutable with each other.
\ts{12pt}Then, there exist the inverses $M^{-1}$ and $\bM^{-1}$, which also satisfy the same properties as $M$, $\bM$. 

\subsection{Construction of Initial Hamitonian System $\mcS$}

\ts{12pt}Consider the dynamical system described by the first-order singular Lagrangian $L$
$$
\begin{array}{rcl}
L&=&L(x,\dot{x},v,\dot{v},\lambda,\dot{\lambda})\vs{12pt}\\

&=&\dot{x}^i\bM_{ij}v_j-
\dps{\f12}\dot{v}_i\Theta^{ij}v_j-\dps{\f12}\dot{x}^i\Xi_{ij}x^j-\dps{\f12}v_iv_i-\lambda \dot{G}(x).
\end{array}
\eqno{(2.5)}
$$
where $x=(x^1,\cdots,x^i\cdots,x^N),v=(v_1,\cdots,v_i,\cdots,v_N)$ and $\dot{G}(x)=\dot{x}^iG_i(x)$\footnote{For any operator $X(x,v)$, in this paper, we shall represent $\partial^x_{i_1}\cdots\partial^x_{i_n}\partial_v^{j_1}\cdots\partial_v^{j_m}X(x,v)$ with $X_{i_1\cdots i_n}^{j_1\cdots j_m}(x,v)$, as far as no ambiguities, where $\partial^x_i=\partial/\partial x^i, \partial_v^i=\partial/\partial v_i$.} \\
\ts{12pt}Following the canonical quantization formulation for constraint systems\cite{A11,A12}, then, the initial unconstraint quantum system $\mcS=(\mcC,\mcA(\mcC),H(\mcC),\mcK)$ is obtained as follows: \vs{12pt}\\
{\bf \boldmath i) Initial canonically conjugate set $\mcC$}
$$ 
\ts{-72pt}\mcC=\{(x^i,p^x_i),(v_i,p_v^i),(\lambda,p_{\lambda})|i=1,\cdots,N\},
\eqno{(2.6)}
$$
which obeys the commutator algebra $\mcA(\mcC)$:
$$
\commut{x^i}{p^x_j}=i\hbar\delta^i_j,\hs{12pt}\commut{v_i}{p_v^j}=i\hbar\delta_i^j,\hs{12pt}
\commut{\lambda}{p_{\lambda}}=i\hbar,\hs{12pt}
\mbox{(the others)}=0,
\eqno{(2.7)}
$$
{\bf \boldmath ii) Initial Hamiltonian $H$}
$$
H=\symmp{\mu^i_{(1)}}{\phi^{(1)}_i}+\symmp{\mu^i_{(2)}}{\phi^{(2)}_i}+\symmp{\mu_{(3)}}{\phi^{(3)}}+\dps{\f12}v_iv_i,
\eqno{(2.8)}
$$
where $\phi^{(n)}$, $(n=1,\cdots, 3)$ are the constraint operators corresponding to the primary constraints $\phi^{(n)}\approx 0$ due to the singularity of the Lagrangian $L$, which are given by
$$
\begin{array}{l}
\phi^{\mbtn{(1)}}_i=\bM_{ij}v_j-p^x_i-\dps{\f12}\Xi_{ij}x^j-\lambda G_i(x),\vs{12pt}\\
\phi^{\mbtn{(2)}}_i=p_v^i+\dps{\f12}\Theta^{ij}v_j,\vs{12pt}\\
\phi^{\mbtn{(3)}}=p_{\lambda},
\end{array}
\eqno{(2.9)}
$$
and $\mu^i_{(1)}$, $\mu^i_{(2)}$ and $\mu_{(3)}$ are the Lagrange multiplier operators.
\vs{12pt}\\
{\bf \boldmath iii) Consistent set of constraints}
\vs{6pt}\\
\ts{12pt}Let $h^{\mbtn{(1)}}_i$, $h^{\mbtn{(2)}}_i$ be
$$
\begin{array}{l}
h^{\mbtn{(1)}}_i=\bM^*_{ij}v_j\vs{12pt}\\
h^{\mbtn{(2)}}_i=-\Xi^*_{ij}v_j,
\end{array}
\eqno{(2.10)}
$$
respectively.\footnote{For any $N\times N$ matrix $A$, $A^*=M^{-1}AM^{{-1}}$. Especially, $\Theta^*_{ij}=M^{-1}\Theta^{ij}M^{-1}.$} From the consistency conditions for the time evolusions of constraint operators, then, the consistent set of constraints, $\mcK$, becomes as follows:
$$
\mcK=\{\phi^{\mbtn{(1)}}_i,\phi^{\mbtn{(2)}}_i,\phi^{\mbtn{(3)}},\psi^{\mbtn{(1)}}\},
\eqno{(2.11)}
$$
where  
$$
\psi^{\mbtn{(1)}}=G_i(x)h^{\mbtn{(1)}}_i=\bM^*_{ij}G_i(x)v_j,
\eqno{(2.12)}
$$
which is the constraint operator corresponding to the secondary constraint. The Lagrange multiplier operators, $\mu^i_{(1)}$, $\mu^i_{(2)}$ and $\mu_{(3)}$ are given by
$$
\begin{array}{l}
\mu_{\mbtn{(1)}}^i=\Theta^*_{ij}G_j(x)\mu_{\mbtn{(3)}}-h^{\mbtn{(1)}}_i,\vs{12pt}\\
\mu_{\mbtn{(2)}}^i=\bM^*_{ij}G_j(x)\mu_{\mbtn{(3)}}-h^{\mbtn{(2)}}_i,\vs{12pt}\\
\mu_{\mbtn{(3)}}=(h^{\mbtn{(1)}}_i\partial^x_i\psi^{\mbtn{(1)}}(x,v)-h^{\mbtn{(2)}}_i\partial^i_v\psi^{\mbtn{(1)}}(x,v))/h^{\mbtn{(3)}},
\end{array}
\eqno{(2.13)}
$$
where
$$
h^{\mbtn{(3)}}=\Theta^*_{ij}(\partial^x_i\psi^{\mbtn{(1)}}(x,v))G_j(x)-\bM^*_{ij}(\partial^v_i\psi^{\mbtn{(1)}}(x,v))G_j(x).
\eqno{(2.14)}
$$
\ts{12pt}The consistent set of constraints, $\mcK$, obeys the commutator algebra $\mcA(\mcK)$:
$$
\begin{array}{l}
\commut{\phi^{\mbtn{(1)}}_i}{\phi^{\mbtn{(1)}}_j}=i\hbar\Xi_{ij},\hs{12pt}\commut{\phi^{\mbtn{(1)}}_i}{\phi^{\mbtn{(2)}}_j}=i\hbar\bM_{ij},\vs{12pt}\\
\commut{\phi^{\mbtn{(2)}}_i}{\phi^{\mbtn{(2)}}_j}=i\hbar\Theta^{ij},\hs{12pt}\commut{\phi^{\mbtn{(1)}}_i}{\phi^{\mbtn{(3)}}}=-i\hbar G_i(x),\vs{12pt}\\
\commut{\phi^{\mbtn{(1)}}_i}{\psi^{\mbtn{(1)}}}=i\hbar G_{ij}(x)\bM^*_{jk}v_k,\vs{12pt}\\
\commut{\phi^{\mbtn{(2)}}_i}{\psi^{\mbtn{(1)}}}=-i\hbar\bM^*_{ij}G_j(x),\hs{12pt}\mbox{(the others)}=0.
\end{array}
\eqno{(2.15)}
$$
\ts{12pt}Thus, we have constructed the initial quamtum system $\mcS$.

\section{The resultant Constraint Quantum System $\mcS^*$}

\ts{12pt}Starting with the initial system $\mcS$, we shall construct the constraint quantum system $\mcS^*$, which satisfies $\mcK=0$, through the star-product quantization formalism of POM\cite{A11}. As well as in our previous work\cite{A2}, this can be accomplished through the successive projections of the initial system. \\
\ts{12pt}For this purpose, we classify $\mcK$ into the following two subsets  :
$$
\mcK=\mcK^{(\mbtn{A})}\oplus\mcK^{\mbtn{(B)}}\q \mbox{with}\q \mcK^{(\mbtn{A})}=\{\phi^{\mbtn{(1)}},\phi^{\mbtn{(2)}}\},\hs{12pt}\mcK^{\mbtn{(B)}}=\{\phi^{\mbtn{(3)}},\psi^{\mbtn{(1)}}\},
\eqno{(3.1)}
$$
and let $\Phat^{(\mbtn{1})}$ be the projection operator associated to the subset $\mcK^{\mbtn{(A)}}$, that is, $\Phat^{\mbtn{(1)}}\mcK^{\mbtn{(A)}}=0$, and $\Phat^{(\mbtn{2})}$, that associated to the subset $\mcK^{\mbtn{(B)}}$. 
From the structure of the commutator algebra (2.15), then, the successive projections of $\mcS$ can be uniquely carried out throuhg the following process: 
$$
\Phat^{\mbtn{(1)}}\mcK^{\mbtn{(A)}}=0\rightarrow\Phat^{\mbtn{(2)}}\mcK^{\mbtn{(B)}}=0,
\eqno{(3.2)}
$$
and, the successive projections of the operators of $\mcS$ are carried out through the following diagram:
$$
\mcC\longrightarrow\mcC^{\mbtn{(1)}}=\Phat^{\mbtn{(1)}}\mcC\longrightarrow\mcC^{\mbtn{(2)}}=\Phat^{\mbtn{(2)}}\mcC^{\mbtn{(1)}}.
\eqno{(3.3)}
$$

\subsection{Projection process $\Phat^{\mbtn{(1)}}\mcK^{\mbtn{(A)}}=0$}

\ts{12pt}Through the projection process $\Phat^{\mbtn{(1)}}\mcK^{\mbtn{(A)}}=0$, the projected system becomes the noncommutative quantum system with
\ts{12pt}Following the POM\cite{A11}, we first construct the ACCS (associated canonically conjugate set) $Z_{\alpha}(=\xi_i(\alpha=i),\e \pi_i(\alpha=i+N);i=1,\cdots,N)$ of the projection operator $\Phat^{\mbtn{(1)}}$. From the commutator algebra $\mcA(\mcK)$, $\{\xi^{\mbtn{(1)}}_i,\pi^{\mbtn{(1)}}_i\}$ satisfy 
$$ 
\begin{array}{l}
\xi_i-\dps{\f12}\Xi_{ij}\pi_j=\phi^{\mbtn{(1)}}_i,\vs{12pt}\\

\pi_i+\dps{\f12}\Theta^{ij}\xi_j=\phi^{\mbtn{(2)}}_i.
\end{array}
\eqno{(3.4)}
$$
Then, $\xi^{\mbtn{(1)}}_i$ and $\pi^{\mbtn{(1)}}_i$ are given as follows:
$$
\begin{array}{l}
\xi^{\mbtn{(1)}}_i=M^{-1}_{ij}(\phi^{\mbtn{(1)}}_j+\dps{\f12}\Xi_{jk}\phi^{\mbtn{(2)}}_k),\vs{12pt}\\
\pi^{\mbtn{(1)}}_i=M^{-1}_{ij}(\phi^{\mbtn{(2)}}_j-\dps{\f12}\Theta^{jk}\phi^{\mbtn{(1)}}_k),
\end{array}
\eqno{(3.5)}
$$
which obey the commutator algebra
$$
\begin{array}{l}
\commut{\xi^{\mbtn{(1)}}_i}{\pi^{\mbtn{(1)}}_j}=i\hbar\delta_{ij},\hs{12pt}\commut{\xi^{\mbtn{(1)}}_i}{\xi^{\mbtn{(1)}}_j}=\commut{\pi^{\mbtn{(1)}}_i}{\pi^{\mbtn{(1)}}_j}=0,\vs{12pt}\\
\commut{Z_{\alpha}}{Z_{\beta}}=i\hbar J^{\alpha\beta}=i\hbar\left(\begin{array}{rr}O&I\\-I&O\end{array}\right)_{\alpha\beta}\hs{6pt}
\mbox{with}\hs{6pt}\left\{\begin{array}{lcl}I&:&N\times N\ \mbox{unit matrix}\\O&:&N\times N\ \mbox{zero matrix}\end{array}
\right.                                                  
\end{array}
\eqno{(3.6)}
$$ 
and satisfy
$$
\Phat^{\mbtn{(1)}}\Phat^{\mbtn{(1)}}=\Phat^{\mbtn{(1)}},\hs{12pt}\Phat^{\mbtn{(1)}} Z_{\alpha}=0,\hs{12pt}\Phat^{\mbtn{(1)}}\Zp_{\alpha}=\Zm_{\alpha}\Phat^{\mbtn{(1)}}=0.
\eqno{(3.7)}
$$
From (2.9), (3.5) and (3.7), $\Phat^{\mbtn{(1)}}$ satisfies the projecion conditions
$$
\begin{array}{l}
\Phat^{\mbtn{(1)}}\phi^{\mbtn{(1)}}_i=\bM_{ij}\Phat^{\mbtn{(1)}}v_i-\Phat^{\mbtn{(1)}}p^x_i-\dps{\f12}\Xi_{ij}\Phat^{\mbtn{(1)}}x^j-\lambda\Phat^{\mbtn{(1)}}G_i(x)=0,\vs{12pt}\\
\Phat^{\mbtn{(1)}}\phi^{\mbtn{(2)}}_i=\Phat^{\mbtn{(1)}}p^i_v+\dps{\f12}\Theta^{ij}\Phat^{\mbtn{(1)}}v_j=0.
\end{array}
\eqno{(3.8)}
$$
\ts{12pt}The {\it hyper}-operator $\Qhat^{\mbtn{(1)}}_{\eta\zeta}$ for $\mcK^{\mbtn{(A)}}$ in the star-product formulation\cite{A11} is given as follows:
$$
\Qhat^{\mbtn{(1)}}_{\eta\zeta}=J^{\alpha\beta}\Zmms{(1)}{\alpha}(\eta)\Zmms{(1)}{\beta}(\zeta)=\ximms{(1)}{i}(\eta)\pimms{(1)}{i}(\zeta)-\pimms{(1)}{i}(\eta)\ximms{(1)}{i}(\zeta),\vs{12pt}\\
\eqno{(3.9)}
$$
The operations of $\ximm{(1)}$ and $\pimm{(1)}$ on $\mcC$ are presented in Appendix A.\\
\ts{12pt}Then, the projected quantum system $\mcS^{\mbtn{(1)}}$ is given by
$$
\mcS^{\mbtn{(1)}}=(\mcC^{\mbtn{(1)}},\mcA(\mcC^{\mbtn{(1)}}),H^{\mbtn{(1)}},\mcK^{\mbtn{(1)}}),
\eqno{(3.10a)}
$$
where
$$
\mcC^{\mbtn{(1)}}=\Phat^{\mbtn{(1)}}\mcC,\hs{6pt}H^{\mbtn{(1)}}=\Phat^{\mbtn{(1)}}H,\hs{6pt}\mcK^{\mbtn{(1)}}=\Phat^{\mbtn{(1)}}\mcK.
\eqno{(3.10b)}
$$

{\subsubsection{Projected CCS $\mcC^{\mbtn{(1)}}$}

Under operation of $\Phat^{\mbtn{(1)}}$ on $\mcS$, the initial CCS, $\mcC$, is projected out to $\mcC^{\mbtn{(1)}}$ as follows: 
$$
\begin{array}{rcl}
\mcC^{\mbtn{(1)}}&=&\mcC\{(\Phat^{\mbtn{(1)}}x,\Phat^{\mbtn{(1)}}p^x),(\Phat^{\mbtn{(1)}}v,\Phat^{\mbtn{(1)}} p_v),(\Phat^{\mbtn{(1)}}\lambda,\Phat^{\mbtn{(1)}}p_{\lambda})\}\vs{6pt}\\
&\equiv&\{x^i,p^x_i,v_i,p^i_v,\lambda,p_{\lambda}|i=1,\cdots,N\},
\end{array}
\eqno{(3.11)}
$$
which satisfies the projection conditions
$$
\begin{array}{l}
\bM_{ij}v_i-p^x_i-\dps{\f12}\Xi_{ij}x^j-\lambda G_i(x)^{\mbtn{(1)}}=0,\vs{12pt}\\
p^i_v+\dps{\f12}\Theta^{ij}v_j=0,
\end{array}
\eqno{(3.12)}
$$
where\footnote{For any operator $F(x,v)(\in \mcC)$, $F(x,v)^{\mbtn{(1)}}=\Phat^{\mbtn{(1)}}F(x,v)$, as far as no  ambiguities.} 
$$
G_i(x)^{\mbtn{(1)}}=\Phat^{\mbtn{(1)}}G_i(x).
\eqno{(3.13)}
$$

\subsubsection{Commutator algebra of $\mcC^{\mbtn{(1)}}$}

\ts{12pt}The commutator algebra of $\mcC^{\mbtn{(1)}}$, $\mcA(\mcC^{\mbtn{(1)}})$, is obtained through the constraint star-product formulation in POM. Taking account of the quantum corrections in the commutator algebra, which are missed in the usual approach with Dirac-bracket quantization formalism, we shall classify $\mcA(\mcC^{\mbtn{(1)}})$ into the following three sub-algebras.\vs{6pt}\\
{\bf Commutator algebra (I)}:
$$
\begin{array}{lcl}

\commut{x^i}{x^j}=i\hbar\Theta^*_{ij},&\hs{12pt}&\commut{x^i}{v^j}=i\hbar\bM^*_{ij},\vs{12pt}\\

\commut{v_i}{v_j}=i\hbar\Xi^*_{ij},& &\commut{\lambda}{p_{\lambda}}=i\hbar,\vs{12pt}\\

\commut{x^i}{p_{\lambda}}=-i\hbar\Theta^*_{ik}G_k(x)^{\mbtn{(1)}},& &\commut{v_i}{p_{\lambda}}=i\hbar\bM^*_{ik}G_k(x)^{\mbtn{(1)}},

\end{array}
\eqno{(3.14a)}
$$
which is constructed with the star-product commutator formulas\cite{A11}. From the projection conditions (3.12), then, the following commutator algebras are obtained.\vs{6pt}\\
{\bf Commutator algebra (II)}:
$$
\begin{array}{l}
\commut{x^i}{p^j_v}=i\hbar\dps{\f12}(\bM\Theta)^*_{ij},\hs{12pt}\commut{v_i}{p^j_v}=i\hbar\dps{\f12}G^*_{ij},\hs{12pt}
\commut{p^i_v}{p^j_v}=-i\hbar\dps{\f14}(G\Theta)^*_{ij},\vs{12pt}\\

\commut{x^i}{p^x_j}=i\hbar(I-\dps{\f12}G^*)_{ij}-i\hbar\lambda\Theta^*_{ik}G_{kj}(x)^{\mbtn{(1)}},\vs{12pt}\\

\commut{v_i}{p^x_j}=i\hbar\dps{\f12}(\Xi\bM)^*_{ij}+i\hbar\lambda\bM^*_{ik}G_{kj}(x)^{\mbtn{(1)}},\vs{12pt}\\

\commut{p^x_i}{p^j_v}=i\hbar\dps{\f14}(\bM G)^*_{ij}-i\hbar\dps{\f12}\lambda G_{ik}(x)^{\mbtn{(1)}}(\bM\Theta)^*_{kj},\vs{12pt}\\

\commut{p^i_v}{p_{\lambda}}=-i\hbar\dps{\f12}(\Theta\bM)^*_{ik}G_{k}(x)^{\mbtn{(1)}}.

\end{array}
\eqno{(3.14b)}
$$
{\bf Commutator algebra (III)}:
\vs{6pt}\\
\ts{12pt}The commutator of $p^x$'s and that between $p^x$ and $p_{\lambda}$ are given, in the following forms, respectively:
$$
\begin{array}{l}
\commut{p^x_i}{p^x_j}=i\hbar C^{{\tiny (p^xp^x)}}_{ij}+i\hbar \mcD^{{\tiny (p^xp^x)}}_{ij},\vs{12pt}\\
\commut{p^x_i}{p_{\lambda}}=i\hbar C^{{\tiny (p^xp_{\lambda})}}_i+i\hbar \mcD^{{\tiny (p^xp_{\lambda})}}_i,
\end{array}
\eqno{(3.14c)}
$$
where
$$
\begin{array}{l}
C^{{\tiny (p^xp^x)}}_{ij}=-\dps{\f14}(G\Xi)^*_{ij}+\dps{\f12}\lambda G_{ik}(x)^{\mbtn{(1)}}G^*_{kj}-\dps{\f12}\lambda G^*_{ik}G_{kj}(x)^{\mbtn{(1)}}
+\lambda^2\Theta^*_{kl}\symmp{G_{ik}(x)^{\mbtn{(1)}}}{G_{jl}(x)^{\mbtn{(1)}}},\vs{12pt}\\
C^{{\tiny (p^xp_{\lambda})}}_i=-\dps{\f12} G^*_{ik}G_{k}(x)^{\mbtn{(1)}}+\lambda\Theta^*_{kl}\symmp{G_{ik}(x)^{\mbtn{(1)}}}{G_l(x)^{\mbtn{(1)}}}.
\end{array}
\eqno{(3.15)}
$$
\ts{12pt}The commutator algebras (I), (II) and the terms $C^{{\tiny (p^xp^x)}}_{ij}, C^{{\tiny (p^xp_{\lambda})}}_i$  in (III) are shown to be equivalent to those due to the Dirac-bracket quantiztion procedure\cite{A12,A13}. On the other hand, the rest terms $\mcD^{{\tiny (p^xp^x)}}_{ij}$, $\mcD^{{\tiny (p^xp_{\lambda})}}_i$ in (III) are the quantum correection terms which are presented in AppendixB. Then, it should be noticed that these correction terms are completely missed in the usual approach with the Dirac-bracket quantization formalism. We shall present the Dirac brackets corresponding to $\mcA(\mcC^{\mbtn{(1)}})$ in Appendix C.

\subsubsection{Projection of Hamiltonian}

\ts{12pt}Taking account of (3.4) and (3.7), the projected Hamiltonian $H^{\mbtn{(1)}}=\Phat^{\mbtn{(1)}}H$ is given in the following form:
$$
H^{\mbtn{(1)}}=\f12v_iv_i+\symmp{(\mu_{\mbtn{(3)}}(x,v))^{\mbtn{(1)}}}{\phi^{\mbtn{(3)}}}+\mcU^{\mbtn{(Q)}},
\eqno{(3.16)}
$$
where $\mcU^{\mbtn{(Q)}}$ is the quantum correction term due to the projection of $\symmp{(\mu_{\mbtn{(3)}}(x,v))}{\phi^{\mbtn{(3)}}}$, which is presented in Appendix B.

\subsubsection{Projection of $\mcK$}

\ts{12pt}From $\Phat^{\mbtn{(1)}}\phi^{\mbtn{(1)}}_i=\Phat^{\mbtn{(1)}}\phi^{\mbtn{(2)}}_i=0$, the projection of $\mcK$ with $\Phat^{\mbtn{(1)}}$ becomes
$$
\begin{array}{lcl}
\mcK^{\mbtn{(1)}}&=&\Phat^{\mbtn{(1)}}\mcK=\Phat^{\mbtn{(1)}}\mcK^{\mbtn{(B)}}=\{\Phat^{\mbtn{(1)}}\phi^{\mbtn{(3)}},\Phat^{\mbtn{(1)}}\psi^{\mbtn{(1)}}\}\vs{12pt}\\
&\equiv&\{\phi^{\mbtn{(3)}},\psi^{\mbtn{(1)}}\},
\end{array}
\eqno{(3.17)}
$$
where
$$
\begin{array}{lcl}
\phi^{\mbtn{(3)}}&=&\Phat^{\mbtn{(1)}}p_{\lambda}=p_{\lambda},\vs{12pt}\\
\psi^{\mbtn{(1)}}&=&\Phat^{\mbtn{(1)}}(\bM^*_{ij}G_i(x)v_j)=\bM^*_{ij}\symmp{G_i(x)^{\mbtn{(1)}}}{v_j}.
\end{array}
\eqno{(3.18)}
$$
Then, the commutator algebra $\mcA(\mcK^{\mbtn{(1)}})$ becomes as follows:
$$
\begin{array}{lcl}
\commut{\psi^{\mbtn{(1)}}}{\phi^{\mbtn{(3)}}}&=&i\hbar(\bM^*)^2_{ij}\symmp{G_i(x)^{\mbtn{(1)}}}{G_j(x)^{\mbtn{(1)}}}-i\hbar\bM^*_{ij}\symmp{\Theta^*_{kl}\symmp{G_{ik}(x)^{\mbtn{(1)}}}{G_l(x)^{\mbtn{(1)}}}}{v_j}\vs{12pt}\\
&-&i\hbar\bM^*_{ij}\symmp{C^{\mbtn{(Q)}}_i(x)}{v_j},
\end{array}
\eqno{(3.19)}
$$
where $C^{\mbtn{(Q)}}_i(x)$ is the quantum correction term completely missed in the usual Dirac-bracket quantiztion, which is presented in Appendix B.

\subsection{Projection process $\Phat^{\mbtn{(2)}}\mcK^{\mbtn{(B)}}=0$}

\ts{12pt}The projection process $\Phat^{\mbtn{(2)}}\mcK^{\mbtn{(B)}}=0$ yields the quantum system $\mcS^*$, which is constrained on the curved space $G(x)=0$.\\
\ts{12pt}Due to the noncommutativity among operators of $x$, there will appear the exremely complicated quantum effect terms in the commutator algebra and operator porducts through the projection process. So, we shall invetigate here the noncommutative quantum system constrained to the curved space within the first order approximation about $\hbar$, that is, the Dirac-bracket quantization procedure, and then, the primary approximation about the noncommutativity-parameters $\theta$, $\eta$\footnote[8]{We shall express these parameters with $\zeta(=\{\theta,\eta\})$ collectively.} will be also taken . 

\subsubsection{Dirac-bracket algebra of $\mcC^{\mbtn{(2)}}$}

\ts{12pt}Let $\varphi^{\mbtn{(1)}}$, $\varphi^{\mbtn{(2)}}$ be
$$
\begin{array}{l}
\varphi^{\mbtn{(1)}}=\psi^{\mbtn{(1)}},\vs{12pt}\\
\varphi^{\mbtn{(2)}}=\phi^{\mbtn{(3)}},
\end{array}
\eqno{(3.20)}
$$
respectively. From (3.19) and (3.20), the Poisson bracket between $\varphi^{\mbtn{(1)}}$ and $\varphi^{\mbtn{(2)}}$ on $\mcC^{\mbtn{(1)}}$ is given as follows:
$$
\commut{\varphi^{\mbtn{(1)}}}{\varphi^{\mbtn{(2)}}}_{\mbtn{PB}}^{\mbtn{(1)}}=\mcW(x)+O(\zeta^2),
\eqno{(3.21)}
$$
where 
$$
\mcW(x)=\mcG(x)(1-\mcG^{-1}(x)\Theta^{ij}G_{ik}(x)G_j(x)v_k)
\eqno{(3.22)}
$$ 
with $\mcG(x)=G_i(x)G_i(x)$. \\
\ts{12pt}According to the iterative property of Dirac bracket\cite{A13}, now, the Dirac bracket for $\mcC^{\mbtn{(2)}}$, $\commut{X}{Y}_{\mbtn{DB}}^{\mbtn{(2)}}$ ($X,Y\in\mcC^{\mbtn{(1)}}$), is defined as follows:
$$
\commut{X}{Y}_{\mbtn{DB}}^{\mbtn{(2)}}=\commut{X}{Y}_{\mbtn{DB}}^{\mbtn{(1)}}-\commut{X}{\varphi^{\mbtn{(n)}}}_{\mbtn{DB}}^{\mbtn{(1)}}\mcW^{-1}(x)\left(\begin{array}{rr}0&-1\\  1&0\end{array}\right)^{(nm)}\commut{\varphi^{\mbtn{(m)}}}{Y}_{\mbtn{DB}}^{\mbtn{(1)}}\hs{12pt}(n,m=1,2),
\eqno{(3.23)}
$$
where $\mcW^{-1}(x)$ is the inverse of $\mcW(x)$,
$$
\mcW^{-1}(x)=\mcG^{-1}(x)(1+\mcG^{-1}(x)\Theta^{ij}G_{ik}(X)G_j(x)v_k)+O(\zeta^2).
\eqno{(3.24)}
$$
Under the Dirac bracket (3.23), the CCS $\mcC^{\mbtn{(1)}}$ is transferred to $\mcC^{\mbtn{(2)}}$,
$$
\begin{array}{rcl}
\mcC^{\mbtn{(2)}}&=&\{x^i,v_i,\lambda|i=1,\cdots,N\}\vs{12pt}\\
\mbox{with}& &p^x_i=v_i-\dps{\f12}\Xi_{ij}x^j-\lambda G_i(x),\vs{12pt}\\
& &p_v^i=-\dps{\f12}\Theta^{ij}v_j,\hs{36pt}p_{\lambda}=0.
\end{array}
\eqno{(3.25)}
$$ 
\ts{12pt}Then, the Dirac-bracket algebra of $\mcC^{\mbtn{(2)}}$, which is denoted by $\mcA_{\mbtn{D}}^{\mbtn{(2)}}$, is represented as follows:

$$
\begin{array}{l}
\commut{x^i}{x^j}_{\mbtn{DB}}^{\mbtn{(2)}}=\Theta^{ij}+D^{(xx)}_{ij},\vs{12pt}\\

\commut{x^i}{v_j}_{\mbtn{DB}}^{\mbtn{(2)}}=P_{ij}(x)+D^{(xv)}_{ij},\vs{12pt}\\
 
\commut{v_i}{v_j}_{\mbtn{DB}}^{\mbtn{(2)}}=\mcG^{-1}(x)(G_{ik}(x)G_j(x)-G_i(x)G_{jk}(x))v_k+\Xi_{ij}+D^{(vv)}_{ij},\vs{12pt}\\

\commut{x^i}{\lambda}_{\mbtn{DB}}^{\mbtn{(2)}}=-\mcG^{-1}(x)G_i(x)+D^{(x\lambda)}_i\vs{12pt}\\

\commut{v_i}{\lambda}_{\mbtn{DB}}^{\mbtn{(2)}}=\mcG^{-1}(x)G_{ij}(x)v_j+D^{(v\lambda)}_i,                          
\end{array}
\eqno{(3.26)}
$$
where
$$
P_{ij}(x)=\delta_{ij}-\mcG^{-1}(x)G_i(x)G_j(x),
\eqno{(3.27)}
$$
$$
\begin{array}{l}
D^{(xx)}_{ij}=\mcG^{-1}(x)(G_i(x)\Theta^{jk}G_k(x)-\Theta^{ik}G_k(x)G_j(x)),\vs{12pt}\\

D^{(xv)}_{ij}=-\mcG^{-2}(x)G_i(x)G_j(x)\Theta^{kl}G_{ks}(x)G_l(x)v_s-\mcG^{-1}(x)\Theta^{ik}(G_{kl}(x)G_j(x)-G_k(x)G_{jl})v_l,\vs{12pt}\\

D^{(vv)}_{ij}=-\mcG^{-1}(x)(\Xi_{ik}G_k(x)G_j(x)-G_i(x)\Xi_{jk}G_k(x))\vs{6pt}\\
\ts{42pt}+\mcG^{-2}(x)(G_{ik}(x)G_j(x)-G_i(x)G_{jk}(x))\Theta^{lm}G_{ln}(x)G_m(x)v_kv_n,\vs{12pt}\\

D^{(x\lambda)}_i=-\mcG^{-1}(x)\Theta^{ij}G_{jk}(x)v_k-\mcG^{-2}(x)G_i(x)\Theta^{jk}G_k(x)G_{jl}(x)v_l,\vs{12pt}\\

D^{(v\lambda)}=-\mcG^{-1}(x)\Xi_{ij}G_j(x)+\mcG^{-2}(x)G_{ij}(x)\Theta^{kl}G_{lm}(x)G_l(x)v_jv_m.
\end{array}
\eqno{(3.28)}
$$
\ts{12pt}Within the 1st-order approximation about $\hbar$ and the noncommutativity- parameters, the commutator algebra for $\mcC^{\mbtn{(2)}}$ is defined as
$$
\commut{X}{Y}^{\mbtn{(2)}}=i\hbar(\commut{X}{Y}_{\mbtn{DB}}^{\mbtn{(2)}})_{\mbtn{(sp)}},
\eqno{(3.29)}
$$
where $(\hs{12pt})_{\mbtn{(sp)}}$ expresses the symmetrized product of $v_i$ with any operator $O(x)$, for examle,
$$
(\mcG^{-2}(x)G_{ij}(x)\Theta^{kl}G_{lm}(x)G_l(x)v_jv_m)_{\mbtn{(sp)}}=\symmp{\mcG^{-2}(x)G_{ij}(x)\Theta^{kl}G_{lm}(x)G_l(x)}{v_jv_m}.
\eqno{(3.30)}
$$
\ts{12pt}The Dirac brackets with respect to $p^x_i$ and $p_v^i$ are estimated from the projection conditions (3.25) and the Dirac-bracket algebra (3.26) in the following way. The Dirac bracket$\commut{x^i}{p^x_j}_{\mbtn{DB}}^{\mbtn{(2)}}$ is obtained as follows:
$$
\begin{array}{lcl}
\commut{x^i}{p^x_j}_{\mbtn{DB}}^{\mbtn{(2)}}&=&\commut{x^i}{v_j-\dps{\f12}\Xi_{jk}x^k-\lambda G_j(x)}_{\mbtn{DB}}^{\mbtn{(2)}}\vs{12pt}\\
&=&\delta^i_j-\lambda\Theta^{ik}G_{kj}(x)+\mcG^{-1}(x)\Theta^{ik}G_k(x)G_{jl}(x)v_l\vs{12pt}\\
& &-\lambda\mcG^{-1}(x)(G_i(x)\Theta^{kl}G_l(x)G_{kj}(x)-\Theta^{ik}G_k(x)G_l(x)G_{lj}(x)).
\end{array}
\eqno{(3.31a)}
$$
Similarly, 
$$
\begin{array}{lcl}
\commut{v_i}{p^x_j}_{\mbtn{DB}}^{\mbtn{(2)}}&=&-\mcG^{-1}(x)G_i(x)G_{jk}v_k+\lambda G_{ij}(x)-\lambda\mcG^{-1}(x)G_i(x)G_k(x)G_{kj}(x)\vs{12pt}\\
&  &+\dps{\f12}\Xi_{ij}+\dps{\f12}\mcG^{-1}(x)G_i(x)\Xi_{jk}G_k(x)-\mcG^{-2}(x)G_i(x)G_{jk}(x)\Theta^{lm}G_{ln}(x)G_mv_nv_k\vs{12pt}\\
& &-\lambda\mcG^{-1}(x)(G_i(x)\Theta^{km}G_{kj}(x)G_{ml}(x)-G_{il}(x)\Theta^{km}G_m(x)G_{kj}(x))v_l\vs{12pt}\\
& &+\lambda\mcG^{-2}(x)G_i(x)G_k(x)G_{kj}(x)G_l(x)\Theta^{lm}G_{ms}(x)v_s,\vs{12pt}\\
\end{array}
\eqno{(3.31b)}
$$
$$
\begin{array}{lcl}
\commut{\lambda}{p^x_j}_{\mbtn{DB}}^{\mbtn{(2)}}&=&-\mcG^{-1}(x)G_{jk}(x)v_k-\lambda\mcG^{-1}(x)G_{jk}(x)\vs{12pt}\\
& &+\dps{\f12}\mcG^{-1}(x)\Xi_{jk}G_k(x)+\mcG^{-2}(x)G_{jk}(x)G_l(x)\Theta^{lm}G_{mn}(x)v_nv_k\vs{12pt}\\
& &-\lambda\mcG^{-1}(x)G_{jk}(x)\Theta^{kl}G_{lm}(x)v_m-\lambda\mcG^{-2}(x)G_{jk}(x)G_k(x)\Theta^{lm}G_m(x)G_{ln}(x)v_n.
\end{array}
\eqno{(3.31c)}
$$
Then, the Dirac bracket $\commut{p^x_i}{p^x_j}_{\mbtn{DB}}^{\mbtn{(2)}}$ is obtained in the following way:
$$
\begin{array}{lcl}
\commut{p^x_i}{p^x_j}_{\mbtn{DB}}^{\mbtn{(2)}}&=&\commut{v_i-\dps{\f12}\Xi_{ik}x^k-\lambda G_i(x)}{p^x_j}_{\mbtn{DB}}^{\mbtn{(2)}}\vs{12pt}\\
&=&\lambda\mcG^{-1}(x)(G_{ik}(x)G_{jl}(x)-G_{jk}(x)G_{il}(x))\Theta^{lm}G_m(x)v_k \vs{12pt}\\
& &+\lambda^2\mcG^{-2}(x)(G_{ik}(x)G_{jl}(x)-G_{jk}(x)G_{il}(x))G_k(x)\Theta^{lm}G_m(x).
\end{array}
\eqno{(3.32)}
$$
From the Dirac brackets (3.26), then, it is shown that the Dirac-bracket algebra $\mcA_{\mbtn{D}}^{\mbtn{(2)}}$ contains the Dirac-bracket algebra of noncommutative system and that on the curved space.\vs{6pt}\\
\ts{12pt}Thus, we have constructed the constraint quantum system 
$$
\mcS^*=(\mcC^*,\mcA(\mcC^*),H^*),
\eqno{(3.33)}
$$
where $\mcC^*=\mcC^{\mbtn{(2)}}$ and
$$
H^*=\f12\symmp{v_i}{v_i}.
\eqno{(3.34)}
$$
Then, it is proved that $\mcS^*$ satisfies the constraint condition $\dot{G}(x)=0$ in the following way:\\
From (3.26),
$$
\commut{G(x)}{v_i}_{\mbtn{DB}}^{\mbtn{(2)}}=G_j(x)\commut{x^j}{v_i}_{\mbtn{DB}}^{\mbtn{(2)}}=0,
\eqno{(3.35)}
$$
therefore
$$
\dot{G}(x)=\commut{G(x)}{H^*}_{\mbtn{DB}}^{\mbtn{(2)}}=\symmp{\commut{G(x)}{v_i}_{\mbtn{DB}}^{\mbtn{(2)}}}{v_i}=0.
\eqno{(3.36)}
$$

\section{Concluding remarks}

\ts{12pt}Starting with the first-order singular Lagrangian containing the term associated to the {\it dynamical} constraint, we have at first constructed the noncommutative quantum system $\mcS^{\mbtn{(1)}}$ in the form including all-orders of the noncommutativity- parameters and $\hbar$. Then, it has been shown that the commutator algebra $\mcA(\mcS^{\mbtn{(1)}})$ and the projected Hamiltonian $H^{\mbtn{(1)}}$ contain the quantum correction terms due to the extreme noncommutativity among the operators $G_i(x)$ associated to the {\it dynamical} constraint, which are completely missed in t he usual approach with the Dirac-baracket quantization. We have next constructed the constraint quantum system $\mcS^*$ within the first-orders of the noncommutativity-parameters and $\hbar$. It has been shown that the commutator algebra $\mcA(\mcC^*)$ with the {\it dynamical} constraint does not conserve the canonically conjugate commutation relations, although those, conserved in the commutative  case\cite{A2}. The exact construction of the noncommutative quantum system on a curved space will be the next task. \vs{48pt}\\

\ts{-18pt}{\bf \Large Appendix}
\appendix
\section{Operations of ACCS(1)}
$$
\begin{array}{lcl}
\ximm{(1)}_kx_i=\inM_{ki},&\hs{12pt}&\pimm{(1)}_kx^i=-\dps{\f12}(\inM\Theta)_{ki},\vs{6pt}\\

\ximm{(1)}_kp^x_i=-\dps{\f12}(\inM\Xi)_{ki}-\lambda \inM_{kl}G_{li}(x),& &\vs{6pt}\\

\pimm{(1)}_kp^x_i=\dps{\f14}(\inM G)_{ki}+\dps{\f12}\lambda(\inM\Theta)_{kl}G_{li}(x),\vs{6pt}\\

\ximm{(1)}_kv_i=-\dps{\f12}(\inM\Xi)_{ki},& &\pimm{(1)}_kv_i=-\inM_{ki},\vs{6pt}\\

\ximm{(1)}_kp_v^i=\inM_{ki},& &\pimm{(1)}_kp^v_i=\dps{\f18}(\inM\Theta G)_{ki},\vs{6pt}\\

\ximm{(1)}_k\lambda=0,& &\pimm{(1)}_k\lambda=0,\vs{6pt}\\

\ximm{(1)}_kp_{\lambda}=-\inM_{kl}G_l(x),& &\pimm{(1)}_kp_{\lambda}=\dps{\f12}(\inM\Theta)_{kl}G_l(x).

\end{array}
\eqno{(\mbox{A}1)}
$$

\newpage
 
\section{Quantum Corrections in Projected System $\mcS^{\mbtn{(1)}}$} 

\subsection{Quantum Corrections in Commutator Algebra $\mcA(\mcC^{\mbtn{(1)}})$}

\ts{12pt}Through the sar-product formulation of POM\cite{A11}, the commutator algebra (III) is represented in the following way: 
$$
\begin{array}{l}
\commut{p^x_i}{p^x_j}=\commut{p^x_i}{p^x_j}^{\mbtn{(1)}}_{\star}\vs{12pt}\\

=-\dps{\f{i\hbar}4}(G\Xi)^*_{ij}+\dps{\f{i\hbar}2}\lambda(G_{ik}(x)^{\mbtn{(1)}}G^*_{kj}- G^*_{ik}G_{kj}(x)^{\mbtn{(1)}})+i\hbar\lambda^2\Theta^*_{kl}\symmp{G_{ik}(x)^{\mbtn{(1)}}}{G_{jl}(x)^{\mbtn{(1)}}}\vs{12pt}\\

+i\hbar\lambda^2\left(\dps{\sum^{\infty}_{n=1}\f{(-1)^n}{(2n+1)!}A_{2n+1}\left(\f{\hbar}2\right)^{2n}(\Theta^*_{k_l})^{2n+1}\symmp{G_{ik_1\cdots k_{2n+1}}(x)^{\mbtn{(1)}}}{G_{jl_1\cdots l_{2n+1}}(x)^{\mbtn{(1)}}}}\right.\vs{12pt}\\

+\left.\dps{\sum^{\infty}_{n=0}\f{(-1)^n}{(2n+2)!}B_{2n+2}\left(\f{\hbar}2\right)^{2n+2}(\Theta^*_{k_l})^{2n+2}(1/i\hbar)\commut{G_{ik_1\cdots k_{2n+2}}(x)^{\mbtn{(1)}}}{G_{jl_1\cdots l_{2n+2}}(x)^{\mbtn{(1)}}}}\right)
\end{array}
\eqno{(\mbox{B}1)}
$$
and
$$
\begin{array}{l}
\commut{p^x_i}{p_{\lambda}}=\commut{p^x_i}{p_{\lambda}}^{\mbtn{(1)}}_{\star}\vs{12pt}\\

=-\dps{\f{i\hbar}2}G^*_{ik}G_{k}(x)^{\mbtn{(1)}}+i\hbar\lambda\Theta^*_{kl}\symmp{G_{ik}(x)^{\mbtn{(1)}}}{G_l(x)^{\mbtn{(1)}}}\vs{12pt}\\

+i\hbar\lambda\left(\dps{\sum^{\infty}_{n=1}\f{(-1)^n}{(2n+1)!}A_{2n+1}\left(\f{\hbar}2\right)^{2n}(\Theta^*_{k_l})^{2n+1}\symmp{G_{ik_1\cdots k_{2n+1}}(x)^{\mbtn{(1)}}}{G_{l_1\cdots l_{2n+1}}(x)^{\mbtn{(1)}}}}\right.\vs{12pt}\\

+\left.\dps{\sum^{\infty}_{n=0}\f{(-1)^n}{(2n+2)!}B_{2n+2}\left(\f{\hbar}2\right)^{2n+2}(\Theta^*_{k_l})^{2n+2}(1/i\hbar)\commut{G_{ik_1\cdots k_{2n+2}}(x)^{\mbtn{(1)}}}{G_{l_1\cdots l_{2n+2}}(x)^{\mbtn{(1)}}}}\right)
\end{array}
\eqno{(\mbox{B}2)}
$$
with $(\Theta^*_{ij})^n=\Theta^*_{{i_1}{j_1}}\cdots\Theta^*_{{i_n}{j_n}}$, where
$$
\dps{A_{2n+1}=\sum^n_{m=0}{}_{2n+1}C_{2m},\hs{36pt}B_{2n+2}=\sum^n_{m=0}{}_{2n+2}C_{2m+1}}
$$
with the binomial coefficient ${}_nC_m=n!/((n-m)!m!)$. \\
\ts{12pt}Thus, the quantum correction terms $\mcD^{{\tiny (p^xp^x)}}_{ij}$ and $\mcD^{{\tiny (p^xp_{\lambda})}}_{ij}$ are given as
$$
\begin{array}{l}
\mcD^{{\tiny (p^xp^x)}}_{ij}=\lambda^2\dps{\sum^{\infty}_{n=1}\f{(-1)^n}{(2n+1)!}A_{2n+1}\left(\f{\hbar}2\right)^{2n}(\Theta^*_{k_l})^{2n+1}\symmp{G_{ik_1\cdots k_{2n+1}}(x)^{\mbtn{(1)}}}{G_{jl_1\cdots l_{2n+1}}(x)^{\mbtn{(1)}}}}\vs{12pt}\\

+\lambda^2\dps{\sum^{\infty}_{n=0}\f{(-1)^n}{(2n+2)!}B_{2n+2}\left(\f{\hbar}2\right)^{2n+2}(\Theta^*_{k_l})^{2n+2}(1/i\hbar)\commut{G_{ik_1\cdots k_{2n+2}}(x)^{\mbtn{(1)}}}{G_{jl_1\cdots l_{2n+2}}(x)^{\mbtn{(1)}}}},
\end{array}
\eqno{(\mbox{B}3)}
$$
$$
\begin{array}{l}
\mcD^{{\tiny (p^xp_{\lambda})}}_i=\lambda\dps{\sum^{\infty}_{n=1}\f{(-1)^n}{(2n+1)!}A_{2n+1}\left(\f{\hbar}2\right)^{2n}(\Theta^*_{k_l})^{2n+1}\symmp{G_{ik_1\cdots k_{2n+1}}(x)^{\mbtn{(1)}}}{G_{l_1\cdots l_{2n+1}}(x)^{\mbtn{(1)}}}}\vs{12pt}\\

+\lambda\dps{\sum^{\infty}_{n=0}\f{(-1)^n}{(2n+2)!}B_{2n+2}\left(\f{\hbar}2\right)^{2n+2}(\Theta^*_{k_l})^{2n+2}(1/i\hbar)\commut{G_{ik_1\cdots k_{2n+2}}(x)^{\mbtn{(1)}}}{G_{l_1\cdots l_{2n+2}}(x)^{\mbtn{(1)}}}}.
\end{array}
\eqno{(\mbox{B}4)}
$$

\subsection{Quantum Corrections in Projected Hamiltonian $H^{\mbtn{(1)}}$} 

\ts{12pt}The projection of $\symmp{\mu_{\mbtn{(3)}}(x,v)}{\phi^{\mbtn{(3)}}}$ in $H$ is obtained with the star-product formulation for the symmetrized product\cite{A11} as follows:
$$
\begin{array}{lcl}
\Phat^{\mbtn{(1)}}\symmp{\mu_{\mbtn{(3)}}(x,v)}{\phi^{\mbtn{(3)}}}&=&\Phat^{\mbtn{(1)}}\symmp{\mu_{\mbtn{(3)}}(x,v)}{p_{\lambda}}=\symmp{\mu_{\mbtn{(3)}}(x,v)}{p_{\lambda}}_{\pscrp\star}\vs{12pt}\\
&=&\symmp{(\mu_{\mbtn{(3)}}(x,v))^{\mbtn{(1)}}}{\phi^{\mbtn{(3)}}}+\mcU^{\mbtn{(Q)}}_{\mbtn{S}}+\mcU^{\mbtn{(Q)}}_{\mbtn{C}},
\end{array}
\eqno{(\mbox{B}5)}
$$   
where
$$
\begin{array}{lcl}
\mcU^{\mbtn{(Q)}}_{\mbtn{S}}=&-&\dps{\sum_{\stackrel{(n+m\neq 0)}{n,m=0}}^{\infty}\f{(-1)^{n+m}}{(2n)!(2m)!}\left(\f{\hbar}2\right)^{2n+2m}(\Theta^*_{ij})^{2n}(\bM^*_{kl})^{2m}}\vs{12pt}\\
& &\times\symmp{(\mu_{\mbtn{(3)}i_1\cdots i_{2n}}^{k_1\cdots k_{2m}}(x,v))^{\mbtn{(1)}}}{G_{j_1\cdots j_{2n}l_1\cdots l_{2m}}(x)^{\mbtn{(1)}}}\vs{18pt}\\

&-&\dps{\left(\f{\hbar}2\right)^2\sum_{n,m=0}^{\infty}\f{(-1)^{n+m}}{(2n+1)!(2m+1)!}\left(\f{\hbar}2\right)^{2n+2m}(\Theta^*_{ij})^{2n+1}(\bM^*_{kl})^{2m+1}}\vs{12pt}\\

& &\times\dps{\symmp{(\mu_{\mbtn{(3)}i_1\cdots i_{2n+1}}^{k_1\cdots k_{2m+1}}(x,v))^{\mbtn{(1)}}}{G_{j_1\cdots j_{2n+1}l_1\cdots l_{2m+1}}(x)^{\mbtn{(1)}}}}
\end{array}
\eqno{(\mbox{B}6)}
$$
and
$$
\begin{array}{lcl}
\mcU^{\mbtn{(Q)}}_{\mbtn{C}}=&-&\dps{\left(\f{\hbar}2\right)^2\sum_{n,m=0}^{\infty}\f{(-1)^{n+m}}{(2n+1)!(2m)!}\left(\f{\hbar}2\right)^{2n+2m}(\Theta^*_{ij})^{2n+1}(\bM^*_{kl})^{2m}}\vs{12pt}\\

& &\times(1/i\hbar)\commut{(\mu_{\mbtn{(3)}i_1\cdots i_{2n+1}}^{k_1\cdots k_{2m}}(x,v))^{\mbtn{(1)}}}{G_{j_1\cdots j_{2n+1}l_1\cdots l_{2m}}(x)^{\mbtn{(1)}}}\vs{18pt}\\

&+&\dps{\left(\f{\hbar}2\right)^2\sum_{n,m=0}^{\infty}\f{(-1)^{n+m}}{(2n)!(2m+1)!}\left(\f{\hbar}2\right)^{2n+2m}(\Theta^*_{ij})^{2n}(\bM^*_{kl})^{2m+1}}\vs{12pt}\\

& &\times(1/i\hbar)\commut{(\mu_{\mbtn{(3)}i_1\cdots i_{2n}}^{k_1\cdots k_{2m+1}}(x,v))^{\mbtn{(1)}}}{G_{j_1\cdots j_{2n}l_1\cdots l_{2m+1}}(x)^{\mbtn{(1)}}}
\end{array}
\eqno{(\mbox{B}7)}
$$
with $(\bM^*_{ij})^n=\bM^*_{{i_1}{j_1}}\cdots\bM^*_{{i_n}{j_n}}$.\vs{6pt}\\
\ts{12pt}Then, the quantum correction term $\mcU^{\mbtn{(Q)}}$ is given with
$$
\mcU^{\mbtn{(Q)}}=\mcU^{\mbtn{(Q)}}_{\mbtn{S}}+\mcU^{\mbtn{(Q)}}_{\mbtn{C}}.
\eqno{(\mbox{B}8)}
$$

\subsection{Quantum Correction in $\mcA(\mcK^{\mbtn{(1)}})$}

\ts{12pt}The commutation relation between $\psi^{\mbtn{(1)}}(=\PhatI\psi^{\mbtn{(1)}})$ and $\phi^{\mbtn{(3)}}(=\PhatI(\phi^{\mbtn{(3)}})$ in $\mcC^{\mbtn{(1)}}$ is obtained in the following way:
$$
\begin{array}{l}
\commut{\psi^{\mbtn{(1)}}}{\phi^{\mbtn{(3)}}}=\commut{\bM^*_{ij}\symmp{G_i(x)^{\mbtn{(1)}}}{v_j}}{p_{\lambda}}=-\commut{\bM^*_{ij}\symmp{G_i(x)^{\mbtn{(1)}}}{v_j}}{G(x)^{\mbtn{(1)}}}\vs{12pt}\\
=-(\commut{\bM^*_{ij}\symmp{G_i(x)}{v_j}}{G(x)}_{\starp})^{\mbtn{(1)}}\vs{12pt}\\
=i\hbar((\bM^*)^2_{ij}G_i(x)^{\mbtn{(1)}}G_j(x)^{\mbtn{(1)}}-\bM^*_{ij}\Theta^*_{kl}\symmp{\symmp{G_{ik}(x)^{\mbtn{(1)}}}{G_l(x)^{\mbtn{(1)}}}}{v_j})\vs{12pt}\\
-i\hbar\bM^*_{ij}\symmp{\dps{\sum^{\infty}_{n=1}\f{(-1)^nA_{2n+1}}{(2n+1)!}\left(\f{\hbar}2\right)^{2n}(\Theta^*_{kl})^{2n+1}\symmp{G_{ik_1\cdots k_{2n+1}}(x)^{\mbtn{(1)}}}{G_{l_1\cdots l_{2n+1}}(x)^{\mbtn{(1)}}}}}{v_j}\vs{12pt}\\
-i\hbar\bM^*_{ij}\symmp{\dps{\left(\f{\hbar}2\right)^2\sum^{\infty}_{n=0}\f{(-1)^nB_{2n+2}}{(2n+2)!}\left(\f{\hbar}2\right)^{2n}(\Theta^*_{kl})^{2n+2}(1/i\hbar)\commut{G_{ik_1\cdots k_{2n+2}}(x)^{\mbtn{(1)}}}{G_{l_1\cdots l_{2n+2}}(x)^{\mbtn{(1)}}}}}{v_j}.
\end{array}
\eqno{(\mbox{B}9)}
$$
Thus, the quantum correction $C^{\mbtn{(Q)}}_i(x)$ is given by
$$
\begin{array}{l}
C^{\mbtn{(Q)}}_i(x)=\dps{\sum^{\infty}_{n=1}\f{(-1)^nA_{2n+1}}{(2n+1)!}\left(\f{\hbar}2\right)^{2n}(\Theta^*_{kl})^{2n+1}\symmp{G_{ik_1\cdots k_{2n+1}}(x)^{\mbtn{(1)}}}{G_{l_1\cdots l_{2n+1}}(x)^{\mbtn{(1)}}}}\vs{12pt}\\
+\dps{\left(\f{\hbar}2\right)^2\sum^{\infty}_{n=0}\f{(-1)^nB_{2n+2}}{(2n+2)!}\left(\f{\hbar}2\right)^{2n}(\Theta^*_{kl})^{2n+2}(1/i\hbar)\commut{G_{ik_1\cdots k_{2n+2}}(x)^{\mbtn{(1)}}}{G_{l_1\cdots l_{2n+2}}(x)^{\mbtn{(1)}}}}.
\end{array}
\eqno{(\mbox{B}10)}
$$

\section{Dirac-bracket algebra corresponding to $\mcA(\mcC^{\mbtn{(1)}})$}

\ts{12pt}In this Appendix, the Dirac-bracket algebra corresponding to the commutator algebra $\mcA(\mcC^{\mbtn{(1)}})$, which we shall represent with $\mcA_{\mbtn{D}}^{\mbtn{(1)}}$, are presented.\\
\ts{12pt}The Poisson brackets about $\mcK^{(\mbtn{A})}$ are
$$
\commut{\phi^{\mbtn{(1)}}_i}{\phi^{\mbtn{(1)}}_j}_{\mbtn{PB}}=\Xi_{ij},\hs{12pt}\commut{\phi^{\mbtn{(1)}}_i}{\phi^{\mbtn{(2)}}_j}_{\mbtn{PB}}=\bM_{ij},\hs{12pt}\commut{\phi^{\mbtn{(2)}}_i}{\phi^{\mbtn{(2)}}_j}_{\mbtn{PB}}=\Theta^{ij},
\eqno{(\mbox{C}1)}
$$
where $\commut{\hs{6pt}}{ }_{\mbtn{PB}}$ is the Poisson bracket defined on $\mcC$.\\
\ts{12pt}Let the matrix $W$ be 
$$
W^{(nm)}_{ij}=
\left(
\begin{array}{rr}
\commut{\phi^{\mbtn{(1)}}_i}{\phi^{\mbtn{(1)}}_j}_{\mbtn{PB}}&\commut{\phi^{\mbtn{(1)}}_i}{\phi^{\mbtn{(2)}}_j}_{\mbtn{PB}}\vs{6pt}\\
\commut{\phi^{\mbtn{(2)}}_i}{\phi^{\mbtn{(1)}}_j}_{\mbtn{PB}}&\commut{\phi^{\mbtn{(2)}}_i}{\phi^{\mbtn{(2)}}_j}_{\mbtn{PB}}
\end{array}
\right)=
\left(
\begin{array}{rr}
\Xi_{ij}&\bM_{ij}\vs{6pt}\\
-\bM_{ij}&\Theta_{ij}
\end{array}
\right)\hs{12pt}(n,m=1,2).
\eqno{(\mbox{C}2)}
$$
Then, the inverse $W^{-1}$ is given by
$$
(W^{-1})^{(nm)}_{ij}=
\left(
\begin{array}{rr}
\Theta^*_{ij}&-\bM^*_{ij}\vs{6pt}\\
\bM^*_{ij}&\Xi^*_{ij}
\end{array}
\right).
\eqno{(\mbox{C}3)}
$$

The Dirac bracket $\commut{X}{Y}_{\mbtn{DB}}^{\mbtn{(1)}}$ corresponding to $\commut{X^{\mbtn{(1)}}}{Y^{\mbtn{(1)}}}$  is defined by 
$$
\commut{X}{Y}_{\mbtn{DB}}^{\mbtn{(1)}}=\commut{X}{Y}_{\mbtn{PB}}-\commut{X}{{\phi^{\mbtn{(n)}}_i}}_{\mbtn{PB}}(W^{-1})^{(nm)}_{ij}\commut{{\phi^{\mbtn{(m)}}_j}}{Y}_{\mbtn{PB}}.
\eqno{(\mbox{C}4)}
$$
Then, the Dirac-bracket algebra $\mcA_{\mbtn{D}}^{\mbtn{(1)}}$ is given with the same manner to $\mcA(\mcC^{\mbtn{(1)}})$ as follows:\vs{6pt}\\
{\bf Dirac-bracket algebra (I)}:
$$
\begin{array}{lcl}

\commut{x^i}{x^j}_{\mbtn{DB}}^{\mbtn{(1)}}=\Theta^*_{ij},&\hs{12pt}&\commut{x^i}{v^j}_{\mbtn{DB}}^{\mbtn{(1)}}=\bM^*_{ij},\vs{12pt}\\

\commut{v_i}{v_j}_{\mbtn{DB}}^{\mbtn{(1)}}=\Xi^*_{ij},& &\commut{\lambda}{p_{\lambda}}_{\mbtn{DB}}^{\mbtn{(1)}}=1,\vs{12pt}\\

\commut{x^i}{p_{\lambda}}_{\mbtn{DB}}^{\mbtn{(1)}}=-\Theta^*_{ik}G_k(x),& &\commut{v_i}{p_{\lambda}}_{\mbtn{DB}}^{\mbtn{(1)}}=\bM^*_{ik}G_k(x),

\end{array}
\eqno{(\mbox{C}5a)}
$$
{\bf Dirac-bracket algebra (II)}:
$$
\begin{array}{l}
\commut{x^i}{p^j_v}_{\mbtn{DB}}^{\mbtn{(1)}}=\dps{\f12}(\bM\Theta)^*_{ij},\hs{12pt}\commut{v_i}{p^j_v}_{\mbtn{DB}}^{\mbtn{(1)}}=\dps{\f12}G^*_{ij},\hs{12pt}
\commut{p^i_v}{p^j_v}_{\mbtn{DB}}^{\mbtn{(1)}}=-\dps{\f14}(G\Theta)^*_{ij},\vs{12pt}\\

\commut{x^i}{p^x_j}_{\mbtn{DB}}^{\mbtn{(1)}}=(I-\dps{\f12}G^*)_{ij}-\lambda\Theta^*_{ik}G_{kj}(x),\hs{12pt}\commut{v_i}{p^x_j}_{\mbtn{DB}}^{\mbtn{(1)}}=\dps{\f12}(\Xi\bM)^*_{ij}+\lambda\bM^*_{ik}G_{kj}(x),\vs{12pt}\\

\commut{p^x_i}{p^j_v}_{\mbtn{DB}}^{\mbtn{(1)}}=\dps{\f14}(\bM G)^*_{ij}-\dps{\f12}\lambda G_{ik}(x)(\bM\Theta)^*_{kj},\hs{12pt}\commut{p^i_v}{p_{\lambda}}_{\mbtn{DB}}^{\mbtn{(1)}}=-\dps{\f12}(\Theta\bM)^*_{ik}G_{k}(x).

\end{array}
\eqno{(\mbox{C}5b)}
$$
{\bf Dirac-bracket algebra (III)}:
$$
\begin{array}{l}
\commut{p^x_i}{p^x_j}_{\mbtn{DB}}^{\mbtn{(1)}}=-\dps{\f14}(G\Xi)^*_{ij}+\dps{\f12}\lambda G_{ik}(x)G^*_{kj}-\dps{\f12}\lambda G^*_{ik}G_{kj}(x)
+\lambda^2\Theta^*_{kl}\symmp{G_{ik}(x)}{G_{jl}(x)},\vs{12pt}\\
\commut{p^x_i}{p_{\lambda}}_{\mbtn{DB}}^{\mbtn{(1)}}=-\dps{\f12} G^*_{ik}G_{k}(x)+\lambda\Theta^*_{kl}\symmp{G_{ik}(x)}{G_l(x))},
\end{array}
\eqno{(\mbox{C}5c)}
$$

\end{document}